\documentclass[journal=jpclcd,manuscript=article]{achemso}

\usepackage{amsmath}
\usepackage{braket} 
\usepackage{amssymb}
\usepackage{amsthm}
\usepackage{geometry}
\usepackage{indentfirst}
\usepackage{fancyhdr}
\usepackage{titletoc}
\usepackage{titlesec}
\usepackage{booktabs}
\usepackage{threeparttable}
\usepackage{multirow}
\usepackage{color}
\usepackage{rotating}
\usepackage[version=4]{mhchem} 
\setkeys{acs}{maxauthors = 0} 

\newcommand{\aprime}{\mathrm{a}^\prime}
\newcommand{\apprime}{\mathrm{a}^{\prime\prime}}
\newcommand{\Aprime}{\mathrm{A}^\prime}
\newcommand{\Apprime}{\mathrm{A}^{\prime\prime}}
\author{Shoukuan Zhao$^{2}$}
\altaffiliation{Contributed equally to this work}
\author{Diandong Tang$^{1}$}
\altaffiliation{Contributed equally to this work}
\author{Xiaoxiao Xiao$^{1}$}
\author{Ruixia Wang$^{2}$}
\author{Qiming Sun$^{3}$}
\author{Zhen Chen$^{2}$}
\author{Xiaoxia Cai$^{2}$}\email{caixx@baqis.ac.cn}
\author{Zhendong Li$^{1}$}\email{zhendongli@bnu.edu.cn}
\author{Haifeng Yu$^{2,4}$}
\author{Wei-Hai Fang$^{1}$}
\affiliation{
\normalsize{$^1${\it
Key Laboratory of Theoretical and Computational Photochemistry, Ministry of Education College of Chemistry, Beijing Normal University, Beijing 100875, China}}\\
\normalsize{$^2${\it
Beijing Academy of Quantum Information Sciences, Beijing 100193, China}}\\
\normalsize{$^3${\it 
Quantum Engine LLC, Washington 98516, US}}\\
\normalsize{$^4${\it
Hefei National Laboratory, Hefei 230088, China}}
}

\title[An \textsf{achemso} demo]
  {Quantum Computation of Conical Intersections on a Programmable Superconducting Quantum Processor}

\begin{document}

\begin{tocentry}
\includegraphics{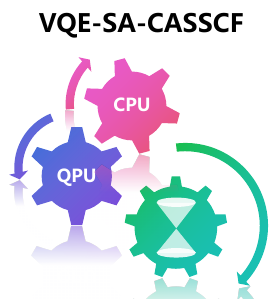}
%
%
%

\end{tocentry}

\begin{abstract}
Conical intersections (CIs) are pivotal in many photochemical processes. Traditional quantum chemistry methods, such as the state-average multi-configurational methods, face computational hurdles in solving the electronic Schr\"{o}dinger equation within the active space on classical computers. While quantum computing offers a potential solution, its feasibility in studying CIs, particularly on real quantum hardware, remains largely unexplored. Here, we present the first successful realization of a hybrid quantum-classical state-average complete active space self-consistent field method based on the variational quantum eigensolver (VQE-SA-CASSCF) on a superconducting quantum processor. This approach is applied to investigate CIs in two prototypical systems - ethylene (\ce{C2H4}) and triatomic hydrogen (\ce{H3}). We illustrate that VQE-SA-CASSCF, coupled with ongoing hardware and algorithmic enhancements, can lead to a correct description of CIs on existing quantum devices. These results lay the groundwork for exploring the potential of quantum computing to study CIs in more complex systems in the future.
\end{abstract}


Conical intersections play a key role in photochemical processes. As the adiabatic potential energy surfaces (PES) of two electronic states at conical intersections become degenerate, ultrafast radiationless transitions from one adiabatic excited state to the ground state or another excited state is possible, which fuels essential photochemical processes such as photoisomerization, photoionization or photocatalysis\cite{domcke2004conical,Worth_2004,Levine_2007,Matsika_2007,Matsika_2011,Levine_2019,Shen_2020,Matsika_2021}. A fundamental task in quantum chemistry is to compute PES accurately, in particular, in the region of conical intersections\cite{Yarkony_1996,Sio_2020}. Since several closely lying electronic states are involved, multi-configurational methods are often needed to correctly describe conical intersections. The standard approach are the state-averaged complete active space self-consistent field (SA-CASSCF)\cite{Lischka_2018} method and its various extensions to include dynamical correlations. The central part of these methods is to solve the electronic Schr\"{o}dinger equation within the active space defined by a set of chemically relevant active orbitals. While there have been successful numerical methods for solving the active space
problem\cite{white1999ab,chan2011density,booth2009fermion}, in the worst scenario the computational cost on classical computers still scales exponentially with respect to the number of active orbitals. Such situation can happen in an exotic nuclear configuration, where conical intersections present.

Quantum computing is generally believed to have the potential to benefit computational physics and quantum chemistry\cite{Aspuru_Guzik_2005,Lanyon_2010,Kassal_2011,Kandala_2017,Cao_2019,McArdle_2020,
bauer2020quantum,motta2021emerging,whitlow2023quantum,valahu2023direct,Wang2023Observation}, as well as promote the development of material science and other related fields. Currently, available quantum hardware usually consists of tens or slightly over one hundred qubits with non-negligible noise, which is often referred to as
the noisy intermediate-scale quantum (NISQ)\cite{preskill2018quantum} device. 
Quantum algorithms based on time evolution, such as quantum phase estimation (QPE)\cite{Malley_2016} and quantum imaginary time evolution (QITE)\cite{motta_2020}, are not feasible for general molecules on current devices due to the significant circuit depth requirement. This motivates the development of hybrid quantum-classical (HQC) algorithms\cite{endo2021hybrid}, such as the variational quantum eigensolver (VQE)\cite{Tilly2022}, which requires shallower circuits and less coherence time, although at the expense of additional classical computational resources. 
Variational HQC algorithms have been developed to compute the ground state of molecules\cite{Peruzzo_2014,Kandala_2017,McClean_2017,Hempel_2018,Grimsley_2019,Jones_2019,Arute_2020,Cerezo_2021,Tang_2021,takeshita2020increasing,tilly2021reduced}, excited states\cite{mcclean2017hybrid,colless2018computation,higgott2019variational,nakanishi2019subspace,parrish2019quantum,Wen_2021,ollitrault2020quantum,wang2024quantum}, molecular response properties\cite{cai2020quantum,chen2021variational,huang2022variational,kumar2023quantum,ziems2024options}, and perform ab initio molecular dynamics\cite{Sokolov_2021,Fedorov_2021}.


Very recently, several efforts have been made to develop quantum algorithms to simultaneously solve the active space problem and optimize molecular orbitals (MOs) 
in either state-specific\cite{sokolov2020quantum,mizukami2020orbital,tilly2021reduced,bierman2023improving,de2023complete}
or state-average formalism\cite{Yalouz_2021,Fitzpatrick_2022,omiya2022analytical,de2023complete}, which are often referred to as (state-average) orbital-optimized VQE (OO-VQE). Besides, algorithms for energy gradients and nonadiabatic
couplings were also developed\cite{omiya2022analytical,Yalouz_2022}, which are essential for performing nonadiabatic molecular dynamics near conical intersections. Despite these algorithmic efforts, it is still very challenging to implement SA-CASSCF based on VQE (VQE-SA-CASSCF) on NISQ devices. One of the major challenge is that unlike VQE for a single
electronic state, SA-CASSCF typically requires to solve the Schr\"{o}dinger equation for multiple electronic states within the active space many times sequentially in order to update the molecular orbitals until convergence. This puts a stringent requirement for the gate fidelity and coherence time of quantum hardware. To the best of our knowledge, despite its importance in the study of nonadiabatic photochemical processes, successful applications of VQE-SA-CASSCF to conical intersections have not yet been reported on real quantum devices.

In this Letter, we report the first successful realization of VQE-SA-CASSCF on a programmable superconducting quantum processor and its applications to conical intersections of two prototypical molecules - ethylene (\ce{C2H4}) and triatomic hydrogen (\ce{H3}). We show that VQE-SA-CASSCF can reproduce conical intersections for these molecules thorough a combination of different strategies, including improving the stability of quantum hardware, reducing the depth of variational circuits, grouping Pauli terms to minimize measurements, and using error mitigation (EM) techniques. 
These results provide a foundation for studying conical intersections of more complex systems using quantum computers in the future.

\begin{figure*}[t]
\centering
\includegraphics[width=0.98\textwidth]{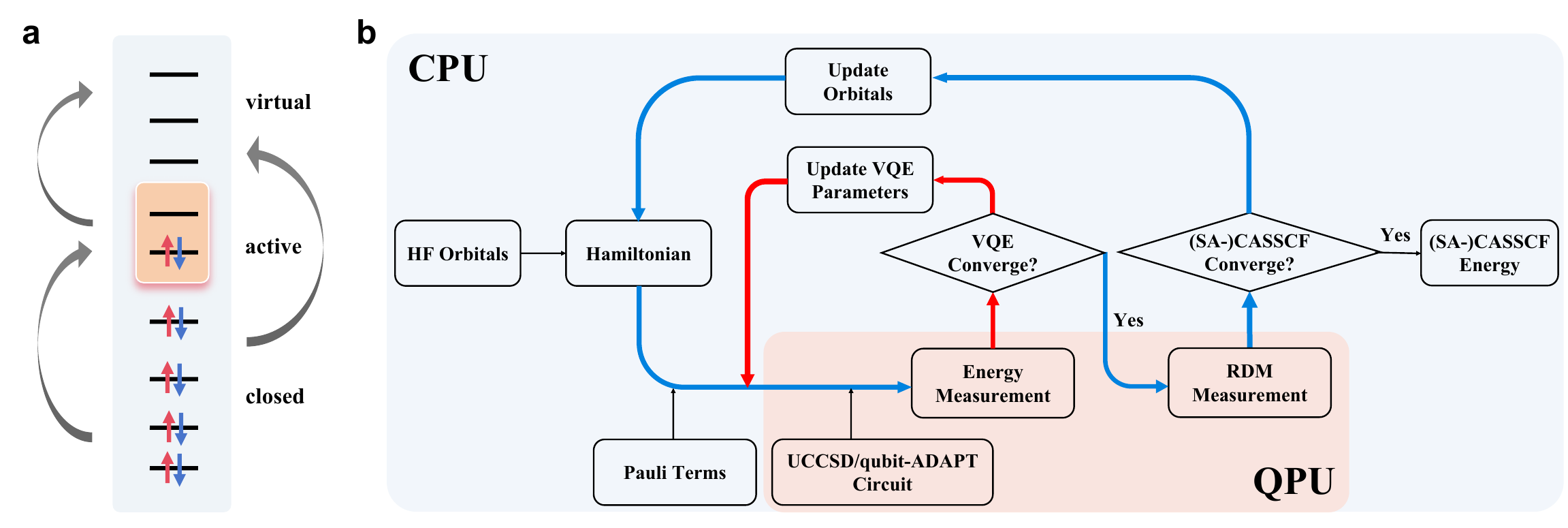}
\caption{(a) Complete active space (CAS) model. The molecular orbitals are partitioned into three classes: closed-shell orbitals with double occupancy, active orbitals with partial occupancy, and virtual orbitals with zero occupancy. The CASSCF method is defined as a variational method, which solves the active space problem exactly while optimizing the orbitals thorough orbital rotations (gray arrows). (b) Flowchart of VQE-SA-CASSCF. The whole procedure involves two closed-loop iterative processes. One is for solving the active space problem using the hybrid quantum-classical VQE (red arrows), and the other is for optimizing molecular orbitals (blue arrows) on classical computers given the reduced density matrices (RDMs) produced by the former part. The single update of both VQE parameters and the molecular orbitals will be referred to as one macro iteration.
}\label{fig:flowchart}
\end{figure*}

{\it VQE-SA-CASSCF method on quantum hardware.}
\noindent For molecules and materials, the electronic Schr\"{o}dinger equation to be solved reads $\hat{H}|\Psi\rangle = E|\Psi\rangle$, where the Hamiltonian $\hat{H}$ in second-quantization is written as\cite{helgaker2014molecular}
\begin{eqnarray}
\hat{H} = \sum_{pq}h_{pq}\hat{a}_p^\dagger \hat{a}_q + \frac{1}{4}\sum_{pqrs}
v_{pq,rs}\hat{a}_p^\dagger \hat{a}_q^\dagger \hat{a}_s \hat{a}_r,\label{eq:Hsq}
\end{eqnarray}
with $h_{pq}$ and $v_{pq,rs}$ being molecular integrals computable on classical computers and $\hat{a}_q^{(\dagger)}$ being Fermionic annihilation (creation) operators. To solve this equation using quantum computers, the problem
needs to be mapped to a qubit problem. There exist different  
fermion-to-qubit transformations such as the Jordan-Wigner\cite{jordan1928pauli},
parity, and Bravyi-Kitaev\cite{Seeley_2012} encodings, after which
$\hat{H}$ becomes a qubit Hamiltonian written as a linear combination of Pauli terms, i.e., $\hat{H}=\sum_k h_k P_k$ with $P_k\in\{I,X,Y,Z\}^{\otimes N}$. 
The many-body wavefunction can be expressed as
$|\Psi\rangle = \sum_{\boldsymbol{q}}\Psi(\boldsymbol{q})|\boldsymbol{q}\rangle$
with $|\boldsymbol{q}\rangle \equiv |q_1\cdots q_N\rangle$ ($q_i\in\{0,1\}$) 
being the occupation number basis vector. 
It is often the case that
only a subset of MOs are relevant for the interested chemical processes, such that it is sensible to only solve Eq. \eqref{eq:Hsq} for
those important orbitals\cite{takeshita2020increasing,tilly2021reduced}. This is the basic idea of CASSCF\cite{Lischka_2018}, in which the MOs are partitioned into three subsets (see Fig. \ref{fig:flowchart}a):
closed-shell orbitals with double occupancy, active orbitals
with partial occupancy, and virtual orbitals with zero occupancy.
The analog of Eq. \eqref{eq:Hsq} is only solved for active orbitals
with the active electrons distributed in all possible ways,
while other parts are treated at a mean-field level.
More concretely, this ansatz is described by the CASCI (complete active space configuration
interaction) wavefunction 
$|\Psi_{\mathrm{CASCI}}^I\rangle=|\Psi_{\mathrm{core}}\rangle|\Psi_{\mathrm{act}}^I\rangle$,
where $|\Psi_{\mathrm{core}}\rangle$ describes the common doubly occupied part
and $|\Psi_{\mathrm{act}}^I\rangle$ describes the correlated many-body wavefunction within the active space for the $I$-th electronic state
($I=1,\cdots,M$ with $M$ being the number of interested electronic states). The CASSCF (complete active space self-consistent field)
ansatz further improves CASCI by allowing orbital rotations among different subspaces\cite{helgaker2014molecular}
\begin{equation}
|\Psi_{\mathrm{CASSCF}}^I\rangle = e^{-\sum_{pq}\kappa_{pq}\hat{a}_{p}^{\dagger}\hat{a}_{q}}
|\Psi_{\mathrm{core}}\rangle|\Psi_{\mathrm{act}}^I\rangle,\label{eq:CASSCF}
\end{equation}
where $\kappa_{pq}$ is an anti-Hermitian matrix for orbital rotations (see Fig. \ref{fig:flowchart}a).
Thus, apart from the wavefunction parameters in $|\Psi_{\mathrm{act}}^I\rangle$,
$\kappa_{pq}$ also needs to be determined by the variational principle.
For simplicity, we will denote these two sets of parameters by
$\mathbf{x}_c^I$ and $\mathbf{x}_o$ for parameters of wavefunctions
and orbitals, respectively.

The flowchart of VQE-SA-CASSCF for optimizing $\mathbf{x}_c^I$ and $\mathbf{x}_o$ used in this work is summarized in Fig. \ref{fig:flowchart}b. The whole procedure involves two closed-loop iterative processes: one
for optimizing $\mathbf{x}_c^I$ using the hybrid quantum-classical VQE and the other for optimizing
$\mathbf{x}_o$ on classical computers. Details of this two-step VQE-SA-CASSCF procedure are described as follows:

(1) Perform a Hartree-Fock calculation to obtain an initial set of MOs.

(2) Construct the active space Hamiltonian using the obtained MOs and
apply a fermion-to-qubit transformation to obtain a qubit Hamiltonian.

(3) Setup a parameterized quantum circuit (PQC) $U_I(\mathbf{x}_c^I)$ 
for each state within the active space $|\Psi^{I}_{\mathrm{act}}(\mathbf{x}_c^I)\rangle\equiv U_I(\mathbf{x}_c^I)|0\rangle$. 
Depending on the complexity of molecules, either the UCCSD (unitary coupled-cluster
singles and doubles)\cite{mcclean2017hybrid}
or qubit-ADAPT (adaptive derivative-assembled problem-tailored)\cite{Grimsley_2019,Tang_2021} ansatz
will be used in this work (vide post).

(4) Solve the active space problem by VQE or its excited-state extensions\cite{higgott2019variational}, which optimizes $\mathbf{x}_c^I$
using both quantum and classical computers. This gives the energy $E_I$ for each state $|\Psi^{I}_{\mathrm{act}}(\mathbf{x}_c^I)\rangle$ as well as the one- and two-particle reduced density matrices (1,2-RDMs) defined by
$\gamma_{pq}^I\equiv\langle\Psi^{I}_{\mathrm{CASSCF}}|\hat{a}_p^\dagger \hat{a}_q|\Psi^{I}_{\mathrm{CASSCF}}\rangle$ 
and $\Gamma_{pqrs}^I\equiv\langle\Psi^{I}_{\mathrm{CASSCF}}|\hat{a}_p^\dagger \hat{a}_q^\dagger \hat{a}_s \hat{a}_r|\Psi^{I}_{\mathrm{CASSCF}}\rangle$, respectively.

(5) With 1,2-RDMs of the $M$ interested electronic states, define a state-average energy function for $\mathbf{x}_o$ as
\begin{eqnarray}
E_{\mathrm{av}}(\mathbf{x}_o)
\equiv 
\sum_{I}
w_I\langle\Psi^I_{\mathrm{CASSCF}}|\hat{H}|
\Psi^I_{\mathrm{CASSCF}}\rangle
=
\sum_{pq}h_{pq}(\mathbf{x}_o)\bar{\gamma}_{pq}
+ \frac{1}{4}\sum_{pqrs}v_{pq,rs}(\mathbf{x}_o)\bar{\Gamma}_{pqrs},\label{eq:Eorb}
\end{eqnarray}
where $\bar{\gamma}_{pq}\equiv\sum_{I=1}^M w_I \gamma_{pq}^{I}$
and $\bar{\Gamma}_{pqrs}\equiv\sum_{I=1}^M w_I \Gamma_{pqrs}^{I}$ with 
$w_I$ being the weight for the $I$-th state 
in SA-CASSCF. Usually, $w_I=1/M$ is chosen.
Then, optimize Eq. \eqref{eq:Eorb} on classical computers
to obtain a set of optimized MOs.

(6) Check energy convergence. If not converged, repeat steps (2)-(6) until convergence
is reached.

\begin{figure*}[t]
\centering
\includegraphics[width=0.98\textwidth]{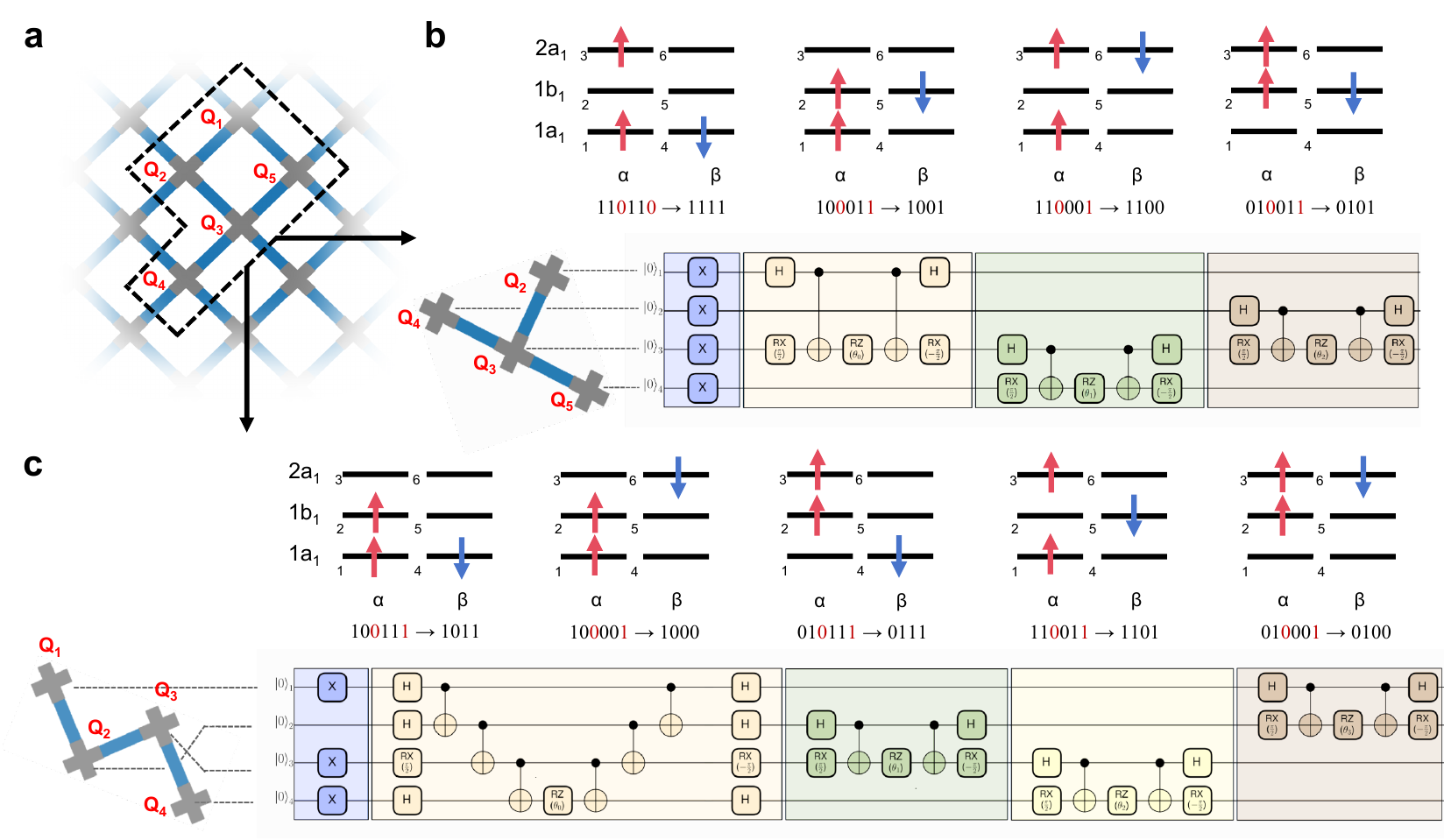}
\caption{
Setup for quantum simulation of the \ce{H3} model using VQE-SA-CASSCF.
(a) Five qubits are picked from the quantum chip to perform the experiments.
(b) Quantum circuit used for the A$_1$
state derived from qubit-ADAPT along with all the involved electronic configurations.
(b) Quantum circuit used for the B$_1$ state.
For each electronic configuration, the bit string on the left side of the arrow represents the corresponding qubit configuration after the parity mapping\cite{Seeley_2012},
while the bit string on the right side represents that after the $Z_2$ reduction\cite{bravyi2017tapering} by removing two bits (red) 
using particle number conservations.}\label{fig:h3}
\end{figure*}

A single pass of steps (2)-(6), which updates both $\mathbf{x}_c^I$ and $\mathbf{x}_o$,
will be referred to as one macro iteration. In step (5), 
we used the second-order approach\cite{Sun_2017} implemented in the PySCF package\cite{sun2018pyscf}.
Our VQE-SA-CASSCF experiments for \ce{C2H4} and \ce{H3} were conducted on a two-dimensional square lattice flip-chip superconducting quantum processor, comprising 63 tunable qubits and 105 tunable couplers\cite{Yan_2018}. Each qubit is connected to 4 adjacent couplers, except for those located at the edge of the lattice,
and each coupler connects two nearest neighbor qubits (see Fig. \ref{fig:h3}a). 
In the experiment for \ce{C2H4}, the qubits Q$_2$ and Q$_3$ were used, while four qubits were used in the experiment for \ce{H3} (see Fig. \ref{fig:h3}).
After careful calibrations, we have achieved an average error rate of less than $1\%$ for the two-qubit CZ gate and around $0.1\%$ for single-qubit gates. For more details about the quantum processor and VQE-SA-CASSCF experiments for \ce{C2H4} and \ce{H4}, see Supplementary information. 


\begin{figure}[t]
\centering
\includegraphics[width=0.8\textwidth]{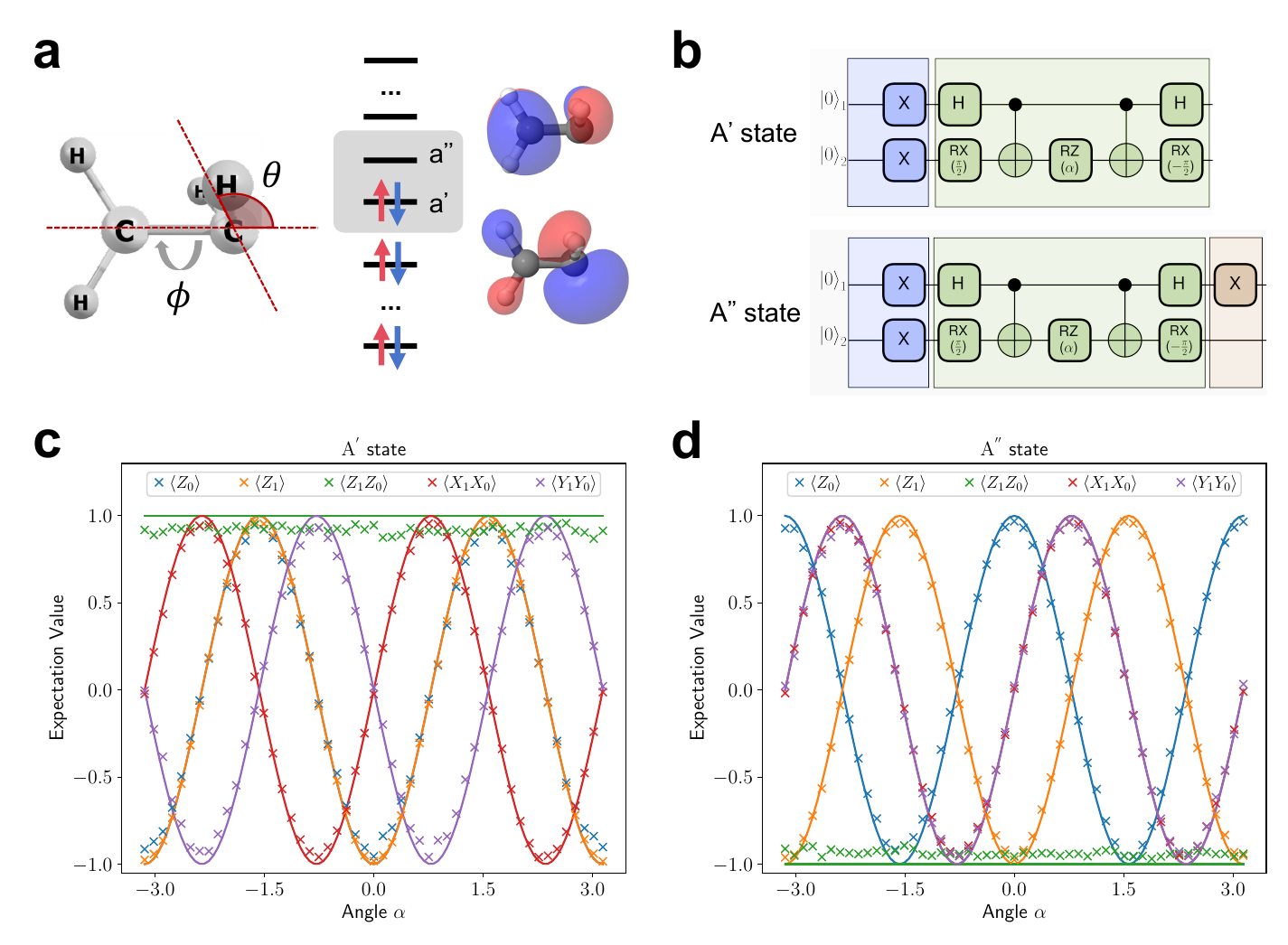}
\caption{(a) Structure of the ethylene \ce{C2H4}, where the torsion angle $\phi$ is fixed at $90^{\circ}$ and the pyramidalization angle $\theta$ is varied to locate the conical intersection between the ground state and the first excited state. The two
active orbitals with $a'$ and $a''$ symmetries in the CAS(2e,2o) model are shown.
(b) The quantum circuits used in the VQE-SA-CASSCF procedure for the $A'$ and $A''$ states. 
(c) Expectation values of the Pauli strings measured from experiments for the $A'$ state. 8000 shots were used for each term. 
(d) Expectation values of the Pauli strings measured from experiments for the $A''$ state.
Solid lines in (c) and (d) represent the theoretical results.
}\label{fig:ethylene}
\end{figure}

\begin{figure}[t]
\centering
\includegraphics[width=0.48\textwidth]{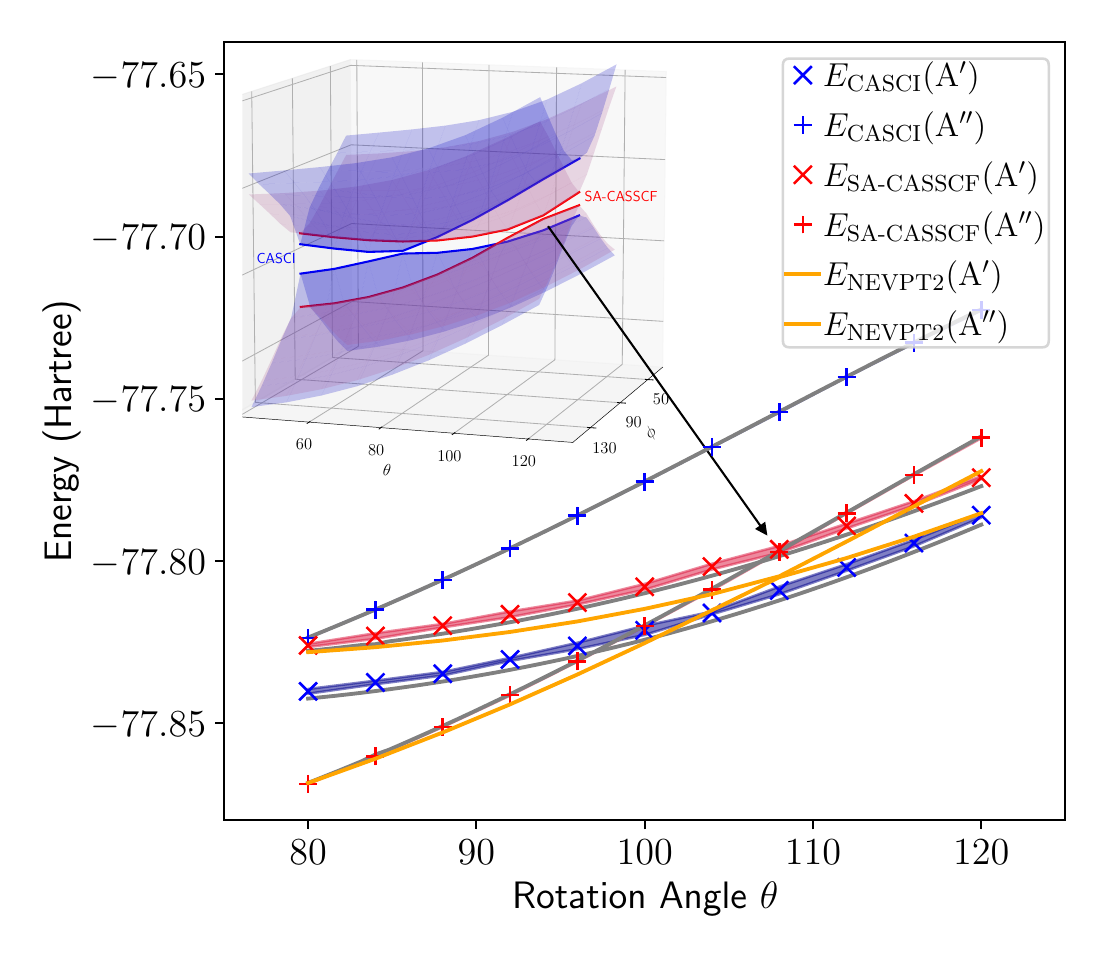}
\caption{
Experimental potential energy curves for the ground and the first excited states of ethylene for different values of $\theta$ but fixed $\phi=90^\circ$ using
VQE-based CASCI and SA-CASSCF with CAS(2e,2o) and the cc-pVDZ basis set.
The CASCI results were obtained by VQE with the Hartree-Fock orbitals, while the SA-CASSCF results were obtained by VQE-SA-CASSCF with the optimized orbitals. The shaded region represents the error bars estimated by repeating the calculations three times. 
For comparison, gray (orange) lines obtained by classical CASCI and SA-CASSCF (NEVPT2) calculations are also shown. The NEVPT2 data are aligned
to the same ground-state energy obtained by SA-CASSCF at $\theta=80^\circ$.
Inset: Conical intersections between the lowest two states obtained by classical CASCI (blue) and SA-CASSCF (red).
}\label{fig:etheneCI}
\end{figure}

{\it Conical intersection in \ce{C2H4}.} We first apply the VQE-SA-CASSCF method to study the conical intersection between the ground and the first excited states of ethylene. Fig. \ref{fig:ethylene}a defines the torsion $\phi$ and pyramidalization $\theta$ degrees of freedom following Ref. \cite{levine2006conical}.~. Specifically, we start from a $C_{2v}$ geometry with the C-C bond fixed at 1.33 \r{A}, all C-H bonds fixed at 1.09 \r{A}, and the H-C-H angles fixed at $116.36^{\circ}$. A conical intersection 
between the lowest two singlet  states ($S_0$ and $S_1$) 
exists at $\phi=90^{\circ}$ and some value of $\theta$ depending on the used method
(see the inset of Fig. \ref{fig:etheneCI}). 
In our experiments, we fixed $\phi=90^{\circ}$ and varied $\theta$ to obtain
two adiabatic potential energy curves (PECs) using VQE-SA-CASSCF
with a two-electron-in-two-orbital active space denoted by CAS(2e,2o),
see Fig. \ref{fig:ethylene}a. Since the geometry at $\phi=90^{\circ}$ possesses a mirror symmetry, the two active orbitals carry the $\aprime$ and $\apprime$ irreducible representations, respectively, and the active parts of the lowest two eigenstates can be expressed by
\begin{eqnarray}
|\Psi_{\Aprime}\rangle  & = & c_1 |\aprime\bar{\aprime}\rangle + c_2 |\apprime\bar{\apprime}\rangle ,\notag\\
|\Psi_{\Apprime}\rangle  & = & d_1 |\aprime\bar{\apprime}\rangle +d_2 |\apprime\bar{\aprime}\rangle,\label{eq:ethyleneWF}
\end{eqnarray}
where $\bar{\aprime}$ represents the $\beta$-spin counterpart.
Since the lowest two eigenstates are of different symmetries at $\phi=90^{\circ}$, VQE can be applied separately by using appropriately designed circuits, which preserve
the spatial symmetry.
After the Bravyi-Kitaev transformation\cite{Seeley_2012} with $Z_{2}$ reduction\cite{bravyi2017tapering}, Eq. \eqref{eq:ethyleneWF} becomes
\begin{eqnarray}
\left|\Psi_{\Aprime}\right\rangle  & = & c_1 \left|11\right\rangle +c_2\left|00\right\rangle ,\notag\\
\left|\Psi_{\Apprime}\right\rangle  & = & d_1 \left|01\right\rangle +d_2 \left|10\right\rangle, 
\end{eqnarray}
which can be described by the PQCs derived from the UCCSD ansatz
shown in Fig. \ref{fig:ethylene}b.

\begin{figure}[t]
\centering
\includegraphics[width=0.8\textwidth]{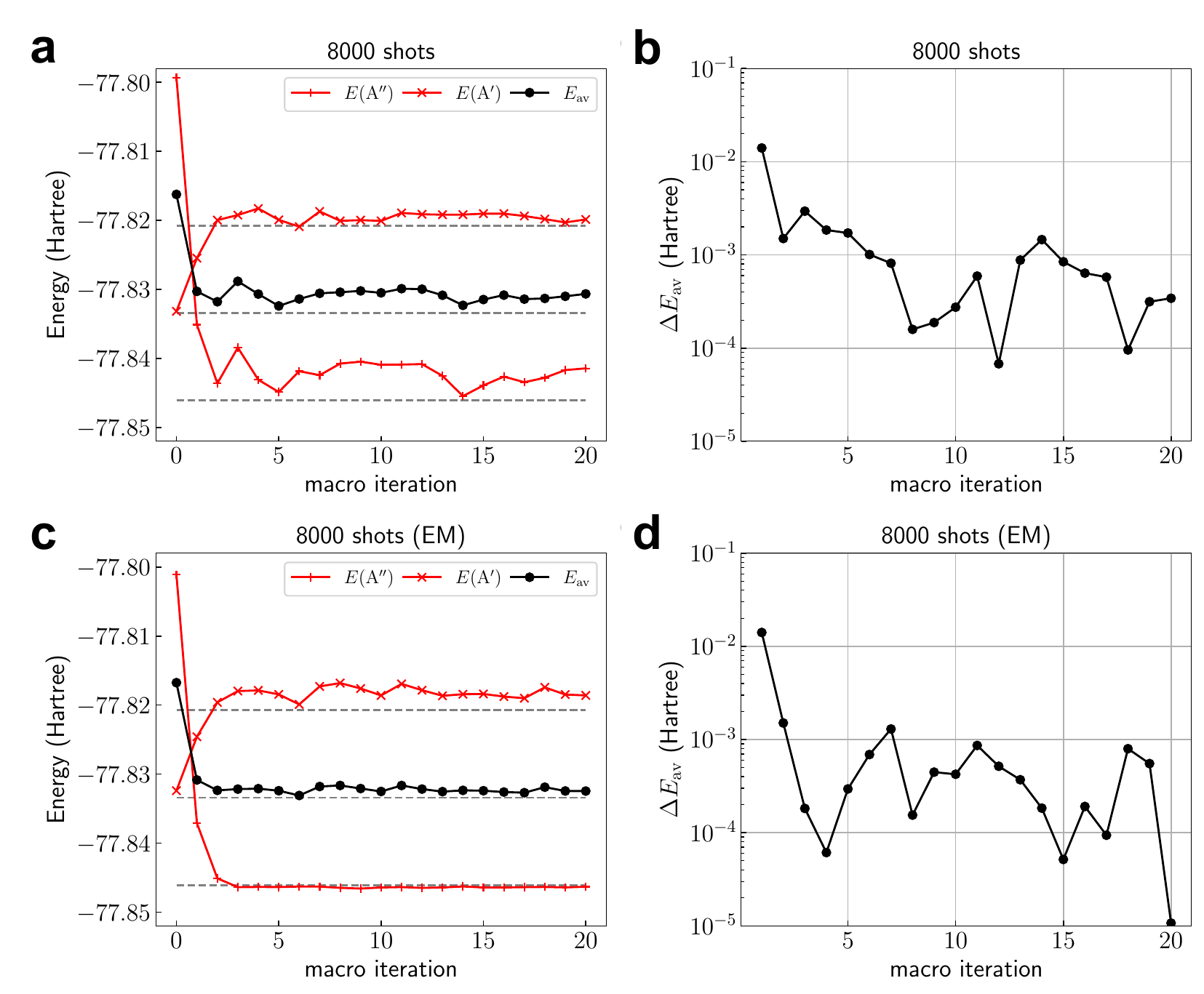}
\caption{Comparison of the optimization processes for VQE-SA-CASSCF without and with error mitigation (EM) for the ethylene model with $\theta=90^{\circ}$
using the cc-pVDZ basis set. (a),(b) VQE-SA-CASSCF without EM, (c),(d) VQE-SA-CASSCF with EM. Gray dashed lines in (a) and (c) represent the converged SA-CASSCF
energies obtained by noiseless numerical simulations. 
(b),(d) the change of the average energy $E_{\mathrm{av}}$ between
two consecutive macro iterations used as the criterion for the convergence
of VQE-SA-CASSCF. For each Pauli term, 8000 shots was used in the measurement
of the expectation value.
}\label{fig:ethyleneEM}
\end{figure}

Figures \ref{fig:ethylene}c and \ref{fig:ethylene}d display the measured expectation values of the Pauli strings used for computing the expectation value
of the Hamiltonian as well as 2-RDM, as a function of the wavefunction parameter $\alpha$. While the experimental expectation values are close to the corresponding theoretical values,
it is still necessary to apply EM strategies for better accuracy.
In this work, we used a simple strategy based on symmetry projection\cite{huang2022variational}, which projects out
nonphysical configurations that violate spatial symmetries (see Methods for details).
The optimization processes for VQE-SA-CASSCF without and with EM for the
ethylene model at $\theta=90^{\circ}$,
using the cc-pVDZ basis set\cite{Dunning1989a} and
the Constrained Optimization BY Linear Approximation (COBYLA) algorithm\cite{Powell1994} for optimizing $\alpha$, are compared in Fig. \ref{fig:ethyleneEM}. Note that each point on
the black curves represents the resulting average energy $E_{\mathrm{av}}$ of two VQE calculations for the two electronic states. It is clear that this EM strategy improves the accuracy of the $\Apprime$ state significantly, leading to a convergence of $E_{\mathrm{av}}$ below $10^{-5}$ Hartree. For the $\Aprime$ state, this EM strategy leads to slightly higher energies, which indicates that the error for this state is not dominated by the error arising
from nonphysical states, but is due to the error within the physical subspace.

With the EM strategy, we performed VQE-SA-CASSCF experiments to study conical intersection of ethylene by varying $\theta$. The PECs for the lowest two
eigenstates are plotted in Fig. \ref{fig:etheneCI}, compared against the ideal results from noiseless simulations (gray lines). 
It is seen that the experimental results are generally 
in good agreements with the theoretical results. 
To demonstrate the benefits of VQE-SA-CASSCF, we also performed
VQE-based CASCI using the Hartree-Fock orbitals. The comparison
in Fig. \ref{fig:etheneCI} reveals that the effects of orbital
optimization are significant, which change the relative ordering
of the two eigenstates for most values of $\theta$ and alter the location
of the conical intersection dramatically. 
Therefore, the more reliable VQE-SA-CASSCF, which gives results close to the more accurate reference data obtained by classical NEVPT2\cite{angeli2001introduction} (second-order n-electron valence state perturbation theory) that further takes into 
account the correlation effects from the orbitals outside the active space, 
is preferred
when the cost of performing multiple VQE calculations are acceptable.
Finally, it is worth noting that without using the CAS model, the original problem would require 96 qubits in the cc-pVDZ basis set and thus huge circuit depths, far beyond the current capabilities of quantum simulation.

\begin{figure}[t]
\centering
\includegraphics[width=0.48\textwidth]{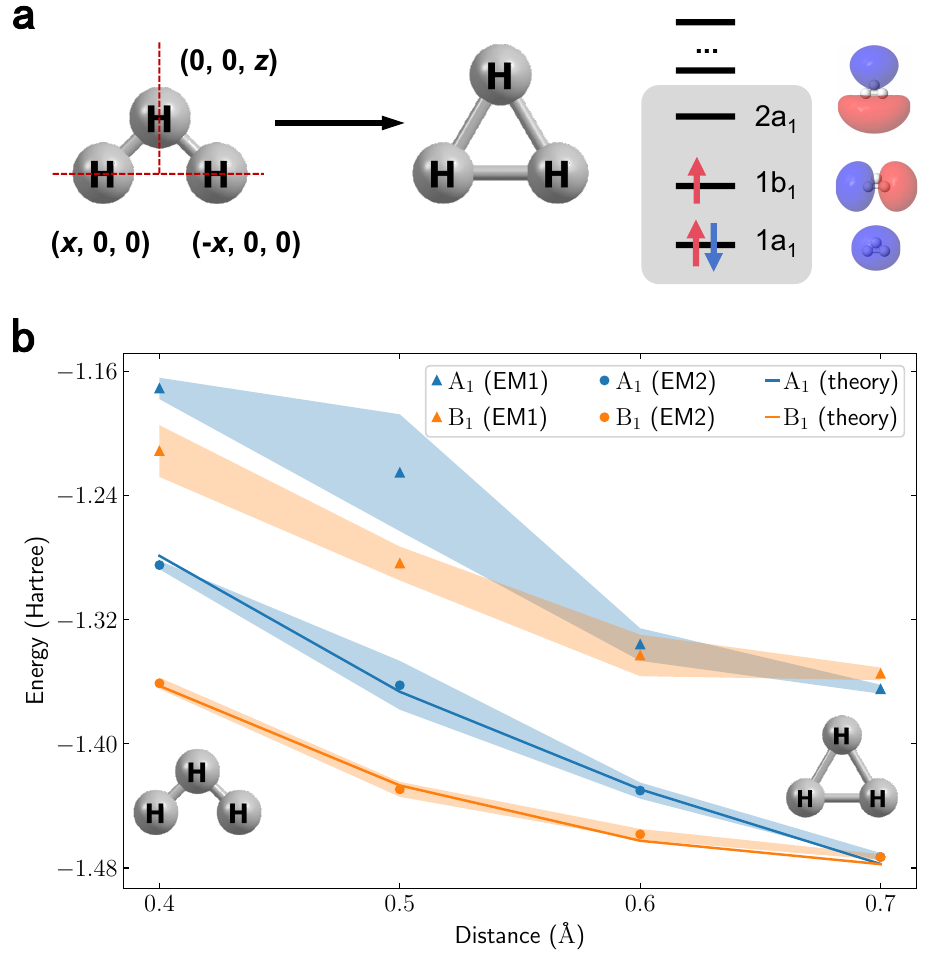}
\caption{(a) Structure of the \ce{H3}
model, where two of the hydrogen atoms 
on the $x$-axis are fixed ($x=0.409$ \AA) and the remaining hydrogen on the $z$-axis
is allowed to move in the $z$ direction.
Molecular orbital diagrams and the three-electron-in-three-orbital active space denoted by CAS(3e,3o). 
(b) Potential energy curves of the ground
and the lowest excited state obtained
by VQE-SA-CASSCF with CAS(3e,3o) using the cc-pVDZ basis set and two error mitigation methods.
The conical intersection is located at the equilateral triangle structure ($z\approx 0.708$ \AA). The shaded region represents the error bars estimated by repeating the
calculations twice.}\label{fig:h3ci}
\end{figure}

{\it Conical intersection in \ce{H3}.} 
The triatomic hydrogen \ce{H3} is another typical system with 
a symmetry-required conical intersection\cite{domcke2004conical}
between the ground and the first excited states
occurred at all equilateral triangular
geometries. In our study, the first two hydrogen atoms are positioned along the $x$-axis with the distance between them set to be 0.818 \r{A},
and the position of the third hydrogen is varied from 0.4 \AA~ to 0.71 \AA~ along the $z$-axis (see Fig. \ref{fig:h3ci}a). A conical intersection between the lowest two electronic states presents at the equilateral triangle structure ($z\approx 0.708$ \AA) with the $D_{3h}$ symmetry, while for other values of $z$, the structure has the $C_{2v}$ symmetry. To correctly describe the conical intersection, a minimal active space with three electrons distributed in three active orbitals, denoted by CAS(3e,3o), is required.  
Figure \ref{fig:h3ci} shows the three relevant MOs with a$_1$, b$_1$, and a$_1$ symmetries, respectively. The ground state and the first excited state has the B$_1$ and A$_1$ symmetries, respectively. The relevant electronic configurations are summarized in Fig. \ref{fig:h3},
along with the qubit configurations obtained after the parity mapping\cite{Seeley_2012}
and $Z_2$ reduction\cite{bravyi2017tapering}.

Realizing VQE-SA-CASSCF for this molecule is much more difficult than the previous case in several aspects. First, the UCCSD ansatz will introduce about 750 quantum gates with around
40 parameters. Second, the number of Pauli terms in the active space Hamiltonian is much
larger than that for the CAS(2e,2o) model of \ce{C2H4}. 
To resolve the first problem, we used the
qubit-ADAPT\cite{Tang_2021} performed on noiseless simulators to derive simpler PQCs (see Methods for details), while maintaining the accuracy with respect to the exact energy below 1 milli-Hartree. The resulting PQCs are summarized in Figs. \ref{fig:h3}a
and \ref{fig:h3}b for the A$_1$ and B$_1$ states, respectively.
Note that the qubits have been properly chosen to avoid nonadjacent two-qubit gates, which would otherwise require the introduction of swap gates. 
To alleviate the second problem, we used a grouping technique\cite{verteletskyi_measurement_2020}, which utilizes the qubit-wise
commutativity between Pauli terms in the Hamiltonian or 2-RDMs,
to reduce the number of measurements
(see Methods for details). 

Apart from these algorithmic improvements, a few other adjustments were implemented to make the VQE-SA-CASSCF experiments robust. On the hardware implementation side, the fast-reset method\cite{McEwen_2021} is utilized to reduce the trigger repeat period to 20 microseconds. Besides, the fidelity of the gate is regularly checked every 10 minutes to mitigate potential fluctuations in the system state. We have also replaced the COBYLA algorithm for optimizing wavefunction parameters by the Bayesian optimization with skopt\cite{scikit_optimize} for having a better performance in the presence of noises. These adjustments are crucial for the VQE-SA-CASSCF experiments.
Note that due the increased impact of noise, VQE-SA-CASSCF can only converge with a loose criteria ($\Delta E_{\mathrm{av}}< 3\times 10^{-3}$ Hartree) in this case. 

Figure \ref{fig:h3ci}b displays the PECs of the ground and the first excited states for \ce{H3} obtained using two different EM strategies. The first EM strategy (labeled by EM1) is the symmetry projection\cite{huang2022variational}, which is the same as that used for \ce{C2H4}. Specifically, the nonphysical states which do not belong to the B$_1$ symmetry for the ground state or the A$_1$ symmetry for the first excited state are projected out.
The second more sophisticated EM strategy (labeled by EM2) further modifies upon the first EM strategy for expectation values of the Pauli-Z operators\cite{newEMmethod2024}, by recycling the information of nonphysical states to suppress the errors from 
depolarization (see Methods for details).
As shown in Fig. \ref{fig:h3ci}b, the PECs obtained by EM1
are too high and have larger fluctuations. In addition, the conical intersection at the equilateral triangular geometry is not well reproduced. 
These suggest that unlike in the two-qubit case with
relatively shorter circuit depth,
simply removing the nonphysical components 
is not sufficient for the present four-qubit case, and the errors within the physical subspace also need to be mitigated.
In comparison, the PECs obtained by EM2 agree much better with the theoretical PECs, reproducing the 
the conical intersection with significantly smaller variations.
Thus, for longer circuits, EM2 is more advantageous than EM1.
This example demonstrates the feasibility of applying near-term quantum computers to study conical intersections using the VQE-SA-CASSCF method
together with hardware and algorithmic improvements.

{\it Conclusion.} This work reports the first successful realization of the quantum-classical hybrid VQE-SA-CASSCF method on quantum computers, and its applications to conical intersections of two prototypical molecules. The most notable difference between VQE-SA-CASSCF and VQE implemented in previous works\cite{Peruzzo_2014,Kandala_2017,Hempel_2018,Arute_2020}
is the simultaneous optimization of wavefunction parameters and 
MOs given the state-average 1,2-RDMs ($\bar{\gamma}_{pq}$ and $\bar{\Gamma}_{pqrs}$) measured on quantum computers. 
While it is crucial for accurately determining the location of conical intersections, 
realizing VQE-SA-CASSCF on near-term quantum devices is often challenging,
because tens or hundreds of VQE calculations need to be carried out sequentially.
This put a stringent requirement for the stability of quantum processors, especially for the superconducting platform with frequency tunable qubits and tunable couplings. 
We demonstrate that a combination of hardware and algorithmic improvements
can successfully alleviate such problem, and lead to a correct reproduction of conical intersections using VQE-SA-CASSCF on NISQ devices. These strategies include using shallower quantum circuits derived from qubit-ADAPT, reducing the number of measurements with grouping techniques, improving the stability of quantum processors, applying sophisticated EM strategies to have more accurate energy functions, etc..


This work raises several interesting points for further study, such as identifying the best alternative quantum solver for the active space problem. One potential solution is to explore the recently developed systematically improvable hardware-efficient ans\"atze\cite{d2023challenges,xiao2023physics}, which may achieve better accuracy with lower circuit depth. Besides, this experiment serves as a reminder that for complicated computational processes involving a large number of iterations, designing better strategies to improve the stability of quantum processors is equally important.

\section*{Methods}
{\it Error mitigation strategies.} The EM strategy based on symmetry projection\cite{huang2022variational} for expectation values in VQE is a generalization of the EM strategy by symmetry verification\cite{bonet2018low,sagastizabal2019experimental}.
Specifically, for a quantum state $|\Psi\rangle$ with certain symmetry and
an operator $\hat{O}$, which commutes with the associated symmetry projector $\mathcal{P}$
(i.e., $\mathcal{P}|\Psi\rangle=|\Psi\rangle$ and $[\hat{O},\mathcal{P}]=0$),
we can use $\langle\Psi|\hat{O}|\Psi\rangle^{\mathrm{EM}}
=\mathrm{tr}(\rho_{\mathrm{exp}}\hat{O}\mathcal{P})/\mathrm{tr}(\rho_{\mathrm{exp}}
\mathcal{P})$ instead of the raw results
$\langle\Psi|\hat{O}|\Psi\rangle^{\mathrm{raw}}\equiv\mathrm{tr}(\rho_{\mathrm{exp}}\hat{O})$
to improve the accuracy of expectation values by removing out contributions
from nonphysical states that violate this symmetry.
For ethylene, the spatial symmetry projectors for the two 
electronic states are formulated as
\begin{eqnarray}
\hat{P}_{\Aprime}  &=& \ket{00}\bra{00} + \ket{11}\bra{11} = \frac{1}{2}(1+Z_1Z_0) \nonumber \\
\hat{P}_{\Apprime} &=& \ket{01}\bra{01} + \ket{10}\bra{10} = \frac{1}{2}(1-Z_1Z_0).
\end{eqnarray}
Then, expectation values of operators such as $X_1X_0$ for the $\Aprime$ state
can be estimated by
\begin{eqnarray}
\langle X_1X_0\rangle^{\mathrm{EM}} &=& 
\frac{\langle X_1X_0\rangle - \langle Y_1Y_0\rangle}{1 + \langle Z_1Z_0\rangle}.
\end{eqnarray}
For \ce{H3}, the symmetry projectors for the A$_1$ and B$_1$ states 
can be derived similarly.
The second EM strategy used for \ce{H3} further modifies the first strategy for expectation values of the Pauli-Z operators by
taking into account the
information of nonphysical states to suppress the errors from 
depolarization\cite{newEMmethod2024}. 
Considering a $n$-layer quantum circuit with global depolarization noise, the density matrix $\rho_{\mathrm{final}}$ is represented by 
\begin{equation}
\rho_{\mathrm{final}} = \frac{P_nI}{d}+(1-P_n)\rho_{\mathrm{ideal}},
\end{equation}
where $P_n=1-(1-p)^n$ with $p$ being the global depolarizing rate for each circuit layer,
$d$ is the dimension of density matrix, and $\rho_{\mathrm{ideal}}$ is the density matrix without noise. For molecules, due to the existence of symmetries such as the conservation of particle number or parity, the ideal density matrix $\rho_{\mathrm{ideal}}$ can be written as 
\begin{eqnarray}
\rho_{\mathrm{ideal}} = \left[\begin{array}{cc}
 \rho_{\mathrm{sym}} & 0 \\
  0 & 0_m
\end{array}\right],
\label{eqM}
\end{eqnarray}
where $\rho_{\mathrm{sym}}$ represents the sub-block with correct symmetries. Then, $\rho_{\mathrm{final}}$ can be rewritten as
\begin{eqnarray}
\rho_{\mathrm{final}} = \left[\begin{array}{cc}
 (1-P_n)\rho_{\mathrm{sym}}+\frac{P_n}{d}I & 0 \\
  0 & \frac{P_n}{d}I_m
\end{array}\right],
\end{eqnarray}
which implies that the ideal probability $p^i_{\mathrm{sym}}$ for the $i$-th physical state can be simply obtained from the final distribution through
\begin{equation}
p_{\mathrm{sym}}^i = \frac{p_{\mathrm{final}}^i-\frac{P_n}{d}}{\sum_i (p_{\mathrm{final}}^i-\frac{P_n}{d})}.
\end{equation}
In actual experiments, $\frac{P_n}{d}$ is estimated by $C\equiv\frac{1}{m}\sum_{j=1}^m p_{\mathrm{nphy}}^j$ where $p_{\mathrm{nphy}}^j$ is the probability for the $j$-th nonphysical states obtained in the computational basis measurement. In addition,
$p_{\mathrm{final}}^i-\frac{P_n}{d}$ is replaced by
$\max(0,p_{\mathrm{phy}}^i-C)$ with $p_{\mathrm{phy}}^i$ being the probability for the
$i$-th physical states.

{\it Qubit-ADAPT ans\"{a}tze for \ce{H3}.} We implemented the qubit-ADAPT method\cite{Tang_2021} using the qubit operator
pool derived from UCCSD excitation operators with the OpenFermion\cite{mcclean2020openfermion} and MindQuantum packages\cite{mq_2021}.
It was applied to \ce{H3} at $z=0.4$ for the 
lowest A$_1$ and B$_1$ states. For the A$_1$ state, 
the optimized qubit-ADAPT ansatz with 1 milli-Hartree accuracy consists of 3 
operators $X_0 Y_2$, $X_2 Y_3$, and $X_1 Y_2$. For the B$_1$ state,
the optimized qubit-ADAPT ansatz consists of 4 operators 
$X_0 X_1 Y_2 X_3$, $X_1 Y_2$, $X_2 Y_3$, and $X_0 Y_1$. The corresponding
circuits are displayed in Fig. \ref{fig:h3}.

{\it Grouping strategy for Hamiltonian and RDMs.} We used the grouping technique\cite{verteletskyi_measurement_2020}, which utilizes the qubit-wise commutativity (QWC), for grouping Pauli terms appeared in the Hamiltonian
and 1,2-RDMs of \ce{H3}, such that Pauli terms in the same group can be measured simultaneously. The grouping procedure implemented using the NetworkX library\cite{Hagberg_2008} in this work is summarized as follows: (1) Obtain all Pauli strings for measurements. (2) Build a graph network with Pauli strings as nodes, and add edges
between two nodes if the corresponding Pauli terms qubit-wise commutes.
(3) Construct complement graph for this network.
(4) Use greedy coloring method to generate a colouring strategy for
the complement graph. Pauli strings with same color will form a subgroup
that can be measured simultaneously. It is worth noting that solutions of greedy coloring
algorithm is not unique. Different results may be obtained in
multiple runs but their differences are small\cite{verteletskyi_measurement_2020}.
The number of nonzero expectation values of Pauli strings 
for the A$_1$ (B$_1$) symmetry is 63 (71), while after grouping 
the number of groups is reduced to 16 (26).

\begin{acknowledgement}
This work was supported by the Innovation Program for Quantum Science and Technology (Grant Nos. 2023ZD0300200 and 2021ZD0301800), National Natural Science Foundation of China (Grants Nos. 21973003, 22288201, 22303005, 11890704 and 92365206), and the Fundamental Research Funds for the Central Universities.
\end{acknowledgement}

\begin{suppinfo}

Details of quantum hardware, experimental setups, and classical optimization methods.

\end{suppinfo}

\bibliography{manu}

\end{document}